# Switching of Co magnetization driven by antiferromagnetic-ferromagnetic phase transition of FeRh alloy in Co/FeRh bilayers

Dróżdż P.[1], Ślęzak M.[1], Matlak K.[1], Matlak B.[1], Freindl K.[2], Korecki J.[1,2], Ślęzak T.[1]

[1]*AGH University of Science and Technology, Faculty of Physics and Applied Computer Science, Kraków, Poland*

[2]*Jerzy Haber Institute of Catalysis and Surface Chemistry PAS, Kraków, Poland*

Abstract:

We show that Co spins in Co/FeRh epitaxial bilayers grown on W(110) switch reversibly between the two orthogonal in-plane directions as the FeRh layer undergoes temperature driven antiferromagnetic-ferromagnetic (AFM-FM) phase transition. Switching of Co magnetization is characterized by a hysteretic behavior owing to a temperature hysteresis of the AFM-FM transition in FeRh. The spin reorientation of Co is driven by the evolution interfacial exchange coupling to FeRh system across the AFM-FM process. Our results provide a new method of writing information purely by a local temperature change.

Control of the spin direction in magnetic nanostructures has become a key objective of investigations in nanomagnetism because it is a mandatory requirement for magnetic recording applications. Typically, in ultrathin ferromagnetic films, the magnetization orientation can be tuned in a process of a spin reorientation transition (SRT), which involves thickness or temperature driven switching of the spontaneous magnetization orientation between two directions [1–3]. In our report, we show a new mechanism for reversible in-plane magnetization switching of ferromagnets in contact with FeRh alloy. An FeRh alloy with an equiatomic composition has revealed a temperature induced first-order magnetic transition from an antiferromagnetic (AFM) to ferromagnetic (FM) state at a transition temperature ($T_T$) 350 K and a second-order transition to a paramagnetic (PM) state at $T_C$ = 675 K [4]. This unique AFM-FM magnetic transition (termed as "metamagnetic"), was first discovered in 1938 [5,6] and is accompanied by volume expansion [7,8], a decrease of resistivity [4], and a large change in entropy [9,10]. Recently, FeRh films exhibiting AFM-FM transitions have attracted considerable attention [11–15], owing to their potential applications in new storage media, such as heat assisted magnetic recording (HAMR) [16,17]. Our experiments show that in Co/FeRh bilayers grown on W(110), spins of the epitaxial Co films witch reversibly between two orthogonal in-plane directions, namely [1-10] and [001], as the FeRh system undergoes an AFM-FM phase transition. Switching of Co magnetization is realized purely by variation of temperature and it is induced by a change of the magnetic coupling between the Co and FeRh spin systems that accompanies the AFM-FM phase transition. The temperature driven switching of the Co spins is characterized by a hysteretic behavior originating from the intrinsic temperature hysteresis of the AFM-FM phase transition in the FeRh system.. The phenomenon reported here goes beyond the HAMR concept consisting in temperature modulation of the coercivity [18] and provides a mechanism for writing information purely by changing the local temperature, which could be realized by laser heating or an electrical current flow.

The ultrathin epitaxial FeRh layer with a thickness of 50 Å was grown on a W(110) substrate by elemental co-deposition at room temperature. The nominal Rh atomic concentration was approximately 54%. The sample was post-annealed at 800 K for 30 min to promote formation of the desired B2 structure [19]. The epitaxial character of the FeRh film was confirmed by high-quality low-energy electron diffraction (LEED) patterns, as shown in Fig. 1a, together with a LEED pattern corresponding to the W(110) substrate (Fig. 1b) and Co overlayer (Fig. 1c).

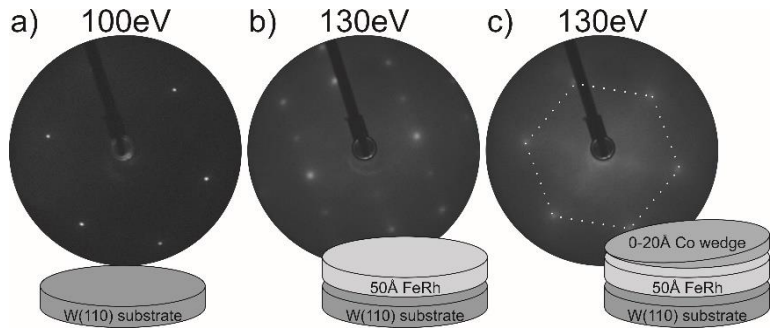

FIG. 1 LEED patterns of a) W(110) single crystal surface, b) FeRh alloy film, and c) cobalt deposited on FeRh system. White dotted line indicate the regular hexagon which marks the positions of diffraction spots. Energy of incident electrons is given above the images.

The magnetic properties were studied in situ by the longitudinal magneto-optical Kerr effect (LMOKE). We used a standard optical layout for the LMOKE setup with s-polarized light ($\lambda = 635$ nm), and photoelastic modulator ($f = 50$ kHz) with lock-in detection. The $2f$ signal measured by the detector was proportional to the Kerr rotation [20]. To obtain the temperature dependence of magnetization, LMOKE magnetic hysteresis loops were measured over a wide temperature range and normalized to the highest value of measured Kerr rotation at saturation $ROT_{sat}$ (corresponding to loop collected at 330 K during cooling process, showed in Fig. 2a). The $ROT_{sat}$ obtained from normalized loops was taken as a measure of the saturation magnetization. The exemplary hysteresis loops recorded with the external magnetic field applied parallel to the [001] in-plane direction of W(110) during the cooling and heating processes together with the corresponding temperature magnetization profile of the AFM-FM phase transition are shown in Fig. 2.

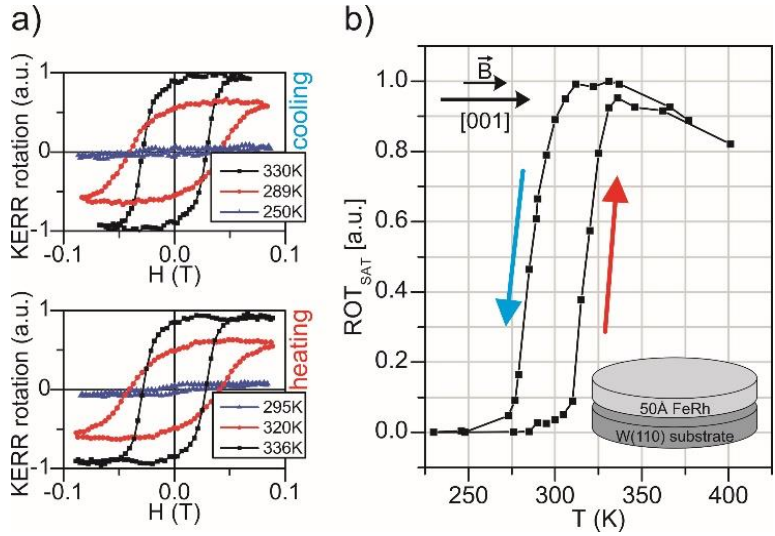

FIG 2. a) The exemplary LMOKE magnetic hysteresis loops acquired across AFM-FM phase transition for 50 Å thick FeRh film on W(110) for the heating and cooling transition branches. b) Temperature magnetization profile of the AFM-FM phase transition.

The critical transition temperatures of the heating and cooling branches were estimated to be $T_{heat}$ = 318 K and $T_{cool}$ = 286 K, respectively, resulting in a transition hysteresis of $\Delta T$ = 32 K. The transition was clearly shifted towards lower temperatures compared with that of the bulk FeRh system ($T_{heat}$ = 360 K, $T_{cool}$ = 345 K [4]) and its hysteresis was broader. Our systematic studies of the thickness dependence of the AFM-FM transition (data not shown) indicated that the major effect on the transition temperature was a lattice strain originated from the considerable mismatch between the W(110) substrate and the FeRh system. This effect allows tailoring of both the critical transition temperature and the hysteresis shape by changing the FeRh film thickness. The in-plane magnetic anisotropy of the discussed FeRh film in the FM was small., as indicated by the similar LMOKE results obtained with an external magnetic field applied along the in plane [1-10] direction (not shown).

On top of the FeRh alloy film, a wedge shaped Co layer with a thickness varying from 0 to 20 Å was grown by MBE at room temperature with the use of a shutter moving in front of the sample during the Co deposition. The deposition of cobalt was followed by annealing at 500 K for 15 min. The LEED pattern corresponding to the Co surface, shown in Fig. 1c with a regular hexagon tailored to the positions of diffraction spots, indicated that Co crystallized on the FeRh(110) plane in a hexagonal hcp phase. Moreover, the LEED patterns measured at specific electron energies were very similar over the whole examined Co thickness range, indicating that the hexagonal symmetry of the Co film surface was preserve over the entire thickness range.

The Co growth was followed by in-situ temperature and Co thickness dependent LMOKE measurements with external magnetic fields applied along the [1-10] and [001] in-plane directions. Figure 3 shows representative hysteresis loops measured for the two distinct Co thicknesses $d_{Co} = 6$ Å and $d_{Co} = 13$ Å at 350 K. This temperature corresponds to the FM state of the FeRh alloy, denoted as $FeRh_{FM}$ and 130 K corresponding to AFM state of FeRh alloy denoted as $FeRh_{AFM}$. The insets in the bottom panel of Fig. 3 show corresponding loops measured at 130 K with an external field applied along the [100] direction.

The hysteresis loops measured at 350 K with an external magnetic field along [1-10] for both thicknesses (top panel of Fig. 3) were rectangular indicating that the easy magnetization direction lays along [1-10]. However, at lower temperatures the magnetic anisotropy clearly changed the orientation of the easy axis to the [001] direction for $d_{Co} = 13$ Å, while the [1-10] direction remained easy for $d_{Co} = 6$ Å. This can be clearly seen from the shape of the loops measured for the [1-10] and [001] directions, respectively (see insets in the bottom panel of Fig. 3). The hard axis loops observed at 130 K for $d_{Co} = 6$ Å along the [001] direction and $d_{Co} = 13$ Å along [1-10] were characterized by a low remanence and high saturation magnetic field. These findings indicate a thickness driven SRT at 130 K.

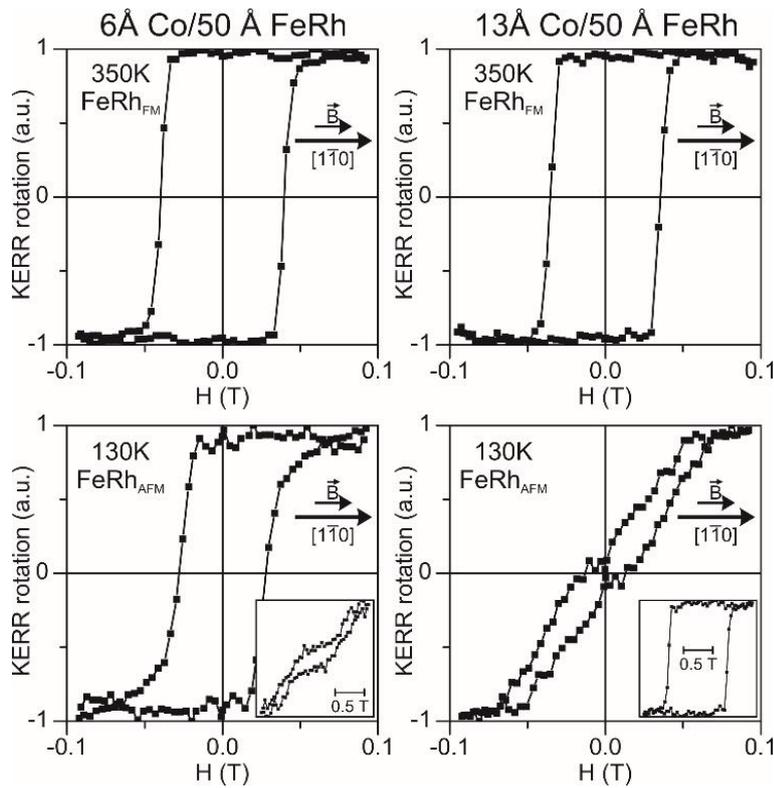

FIG. 3 LMOKE loops measured for the Co/FeRh bilayer with an external magnetic field applied along the [1-10] in-plane direction, shown for cobalt film thicknesses of 6 Å (left panel) and 13 Å (right panel). The top and bottom panels correspond to measurements at 350 and 130 K, respectively. Insets in the bottom panels show LMOKE loops measured along the [001] direction. Loops were normalized to Kerr rotation signal at saturation for each loops.

The systematic LMOKE measurements performed as a function of Co thickness at 130 and 350 K allowed us to derive the dependence of the remanence Kerr signal ROT$_{rem}$ as function of Co thickness, as shown in the Fig. 4a for the discussed temperatures. At 350 K the saturation normalized remanence signal derived from the [1-10] loops was only weakly temperature dependent and close to unity. Additionally, the remanence signal for the [001] direction was nearly zero at all cobalt thicknesses (data not shown). This finding indicates that the easy magnetization axis of the Co/FeRh system at 350 K was along [1-10] over the whole investigated thickness range. In contrast to the 350 K behavior, at 130 K the remanence, derived from the [1-10] loops, decreased from nearly one at a low thickness to zero for cobalt thicker than $d_{crit}$ = 8 Å. This behavior was accompanied by an inverse change of the remanence derived from LMOKE loops along the [001] direction indicating that at low temperatures, corresponding to the AFM state of FeRh the cobalt layer underwent a thickness driven in-plane SRT. Moreover, a comparison of the MOKE remanence thickness dependence between the two discussed temperatures clearly indicated that cobalt thicker than $d_{crit}$= 8 Å featured a temperature driven SRT between the [1-10] direction at 350 K to [001] at 130 K.

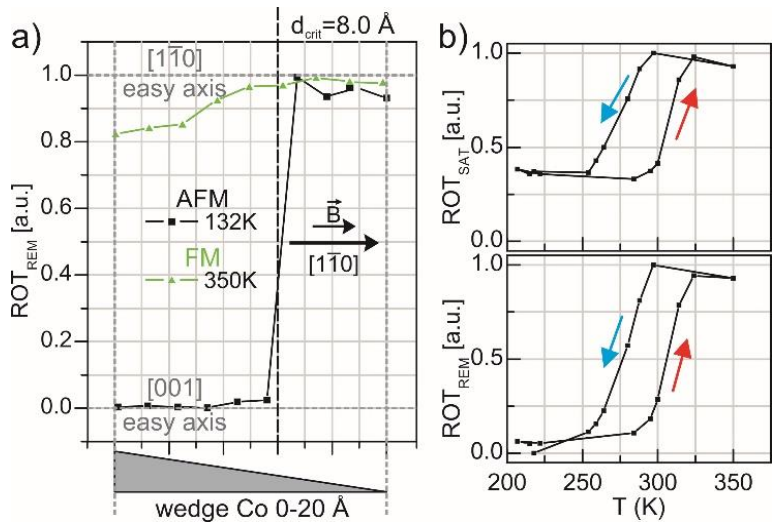

FIG. 4 a) Remanence Kerr rotation ROT$_{REM}$ normalized to saturation determined from LMOKE loops measured with external magnetic field applied along the [1-10] direction, shown as a function of cobalt thickness. b) Temperature dependence of the remanence Kerr rotation ROT$_{REM}$ and saturation Kerr rotation ROT$_{SAT}$ determined from LMOKE loops measured with external magnetic field applied along the [1-10] direction, shown as a function of temperature for $d_{Co}$ = 13 Å.

To understand the origin of the observed SRT we performed temperature dependent LMOKE measurement for a selected Co thickness in the range above $d_{crit}$ of thickness driven SRT at 130 K, namely for $d_{co}$ = 13 Å. The LMOKE loops were measured with an external field applied along [1-10] during the temperature cycle 350 K $\Rightarrow$ 200 K $\Rightarrow$ 350 K. The amplitude of the LMOKE signal corresponding to the saturation state (Fig. 4b top panel) and the normalized remanence Kerr rotation Fig. 4b bottom panel) were determined as a function of temperature.

The temperature evolution of the saturation Kerr signal corresponds to the AFM-FM phase transition in the FeRh and is characterized by a typical transition temperature hysteresis. The critical transition temperatures on heating and cooling branches were estimated to be $T_{heat}$ = 308 K and $T_{cool}$ = 274 K, respectively, resulting in a transition hysteresis $\Delta T$ = 34 K. The transition was shifted by approximately 10 K towards lower temperatures compared with that of the uncoated FeRh film. This shift should be attributed to the exchange coupling with the ferromagnetic cobalt film. In parallel, the remanence Kerr signal is a fingerprint of the magnetization orientation and its temperature induced evolution is a manifestation of the cobalt SRT between the [1-10] direction (high remanence value) and [001] direction (low remanence state). It is clear from a comparison of the bottom and top panel of Fig. 4b that the temperature driven SRT was characterized by an identical hysteresis to that of the AFM-FM phase transition. Thus, we consider that the SRT was induced by the phase transition of the alloy film. This observation shows that by changing the temperature of the Co/FeRh bilayer, the 90-degree magnetization switching can be realized without any additional external factor, as an external magnetic field or electric current. With this effect one can imagine writing information bits, represented by Co magnetization orientation, purely by a small temperature change (for example by a low power laser beam heating). Furthermore, owing to the hysteretic behavior of the cobalt SRT process depending on the sample temperature history at a given temperature near room temperature, both the [1-10] and [001] orientations of cobalt magnetization can be stabilized. Our interpretation of the observed SRT involves the intrinsic uniaxial anisotropy of the FeRh /Co interface with an easy axis parallel to [1-10]. This conclusion is supported by LMOKE loops measured for the whole investigated Co thickness range at temperatures above $T_{AFM-FM}$ and at low temperatures for the low cobalt thickness regime. Furthermore, along with the AFM-FM phase transition, a change between a frustrated and collinear interlayer exchange coupling Co-FeRh took place for the $FeRh_{AFM}$ and $FeRh_{FM}$ states, respectively. In the $FeRh_{AFM}$ state, the cobalt spins cannot simultaneously satisfy the parallel alignment to the opposite Fe spin sublattices of the compensated (110) FeRh plane, which leads to the spin-flop phenomena. This frustration results in an abrupt switching of the Co magnetization to the [001] direction, which is perpendicular to the Fe spins in a FeRh system. Consequently, a lack of an exchange bias was observed in our LMOKE data in agreement with the findings of Schulthess et al. [21] and Moran at al. for the $FeF_2$/Fe system [22]. A schematic drawing of the evolution of the interfacial spin structure across the FM-AFM transition for FeRh covered by Co is shown in Fig. 5.

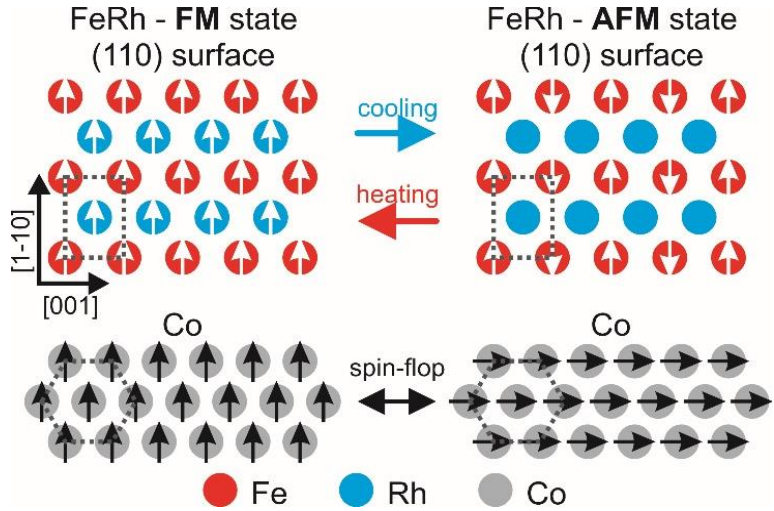

FIG. 5 Schematic showing evolution of interfacial spin structure across FM-AFM transition for Co/FeRh bilayer for $d_{Co}$ = 13 Å. Left and right panels correspond to the FeRh$_{FM}$ and FeRh$_{AFM}$ states of the alloy film, respectively. White and black arrows label the FeRh (red and blue spheres) and Co (gray spheres) spins, respectively.

Finally, the origin of the thickness induced cobalt SRT at low temperatures can be interpreted in a similar way. In the low thickness limit, interfacial contributions preferring the [1-10] magnetization direction are stronger than the spin flop induced uniaxial magnetic anisotropy. Thus, the magnetization aligns parallel to the [1-10] direction. As the thickness increases, the two anisotropic contributions compete. However, the relative impact of the FeRh/Co interface anisotropy decreases with increasing cobalt thickness and above $d_{crit}$ the spin-flop induced anisotropy dominates, forcing the reorientation transition to [001] direction.

In conclusion, we have shown that it is possible to write an information bit (represented by the Co magnetization orientation) in FeRh/Co bilayers by a temperature change alone. Such simple magnetization switching is possible owing to a temperature driven AFM-FM phase transition in the FeRh layer. Our observations indicate a possibility of using the AFM-FM phase transitions in a device (heat engine) that transfers the thermal energy to the magnetic anisotropy energy of a ferromagnetic system neighboring the FeRh layer. Integration of a FeRh system with a properly designed sandwich-like system might result in a manipulation of the magnetization direction of ferromagnetic sublayers in a multilayer stack or even a whole multilayer.

Acknowledgment:

This work was supported by the National Science Center Poland (NCN) under Project No. 2015/19/B/ST3/00543 and scholarship founded by Krakow Consortium "Matter-Energy-Future" the name of Marian Smoluchowski under grant KNOW.

[1] Y. Millev and J. Kirschner, Phys. Rev. B **54**, 4137 (1996).

[2] H. P. Oepen, M. Speckmann, Y. Millev, and J. Kirschner, Phys. Rev. B **55**, 2752 (1997).

[3] S. Hope, E. Gu, B. Choi, and J. A. C. Bland, Phys. Rev. Lett. **80**, 1750 (1998).

[4] J. S. Kouvel and C. C. Hartelius, J. Appl. Phys. **33**, 1343 (1962).

[5] M. Fallot, Ann. Phys. **10**, 291 (1938).

[6] M. Fallot and R. Hocart, Rev. Sci. **77**, 498 (1939).

[7] A. I. Zakharov, A. M. Kadomtseva, R. Z. Levitin, and E. G. Ponyatovskii, Sov. Phys. JEPT **19**, 1348 (1964).

[8] M. R. Ibarra and P. A. Algarabel, Phys. Rev. B **50**, 4196 (1994).

[9] J. S. Kouvel, J. Appl. Phys. **37**, 1257 (1966).

[10] M. P. Annaorazov, S. A. Nikitin, A. L. Tyurin, K. A. Asatryan, and A. K. Dovletov, J. Appl. Phys. **79**, 1689 (1996).

[11] R. O. Cherifi, V. Ivanovskaya, L. C. Phillips, A. Zobelli, I. C. Infante, E. Jacquet, V. Garcia, S. Fusil, P. R. Briddon, N. Guiblin, A. Mougin, A. A. Ünal, F. Kronast, S. Valencia, B. Dkhil, A. Barthélémy, and M. Bibes, Nat. Mater. **13**, 345 (2014).

[12] X. Marti, I. Fina, C. Frontera, J. Liu, P. Wadley, Q. He, R. J. Paull, J. D. Clarkson, J. Kudrnovský, I. Turek, J. Kuneš, D. Yi, J. Chu, C. T. Nelson, L. You, E. Arenholz, S. Salahuddin, J. Fontcuberta, T. Jungwirth, and R. Ramesh, Nat. Mater. **13**, 367 (2014).

[13] C. Bordel, J. Juraszek, D. W. Cooke, C. Baldasseroni, S. Mankovsky, and J. Minar, Phys. Rev. Lett. **109**, 117201 (2012).

[14] Y. Lee, Z. Q. Liu, J. T. Heron, J. D. Clarkson, J. Hong, C. Ko, M. D. Biegalski, U. Aschauer, S. L. Hsu, M. E. Nowakowski, J. Wu, H. M. Christen, S. Salahuddin, J. B. Bokor, N. A. Spaldin, D. G. Schlom, and R. Ramesh, Nat. Commun. Suppl. Mater. **6**, 6959 (2015).

[15] C. Stamm, J. U. Thiele, T. Kachel, I. Radu, P. Ramm, M. Kosuth, J. Minár, H. Ebert, H. A. Dürr, W. Eberhardt, and C. H. Back, Phys. Rev. B **77**, 184401 (2008).

[16] J. Thiele, S. Maat, and E. E. Fullerton, Appl. Phys. Lett. **82**, 2859 (2003).

[17] M. H. Kryder, E. C. Gage, T. W. Mcdaniel, W. a. Challener, R. E. Rottmayer, G. Ju, Y. T. Hsia, and M. F. Erden, Proc. IEEE **96**, 1810 (2008).

[18] N. T. Nam, W. Lu, and T. Suzuki, J. Appl. Phys. **105**, 07D708 (2009).

[19] S. K. Kim, Y. Tian, F. Jona, and P. M. Marcus, Phys. Rev. B **56**, 9858 (1997).

[20] P. Vavassori, Appl. Phys. Lett. **77**, 1605 (2000).

[21] T. Schulthess and W. Butler, Phys. Rev. Lett. **81**, 4516 (1998).

[22] T. J. Moran, J. Nogués, D. Lederman, I. K. Schuller, D. Lederman, and I. K. Schuller, Appl. Phys. Lett. **72**, 1 (1998).